\documentclass[11pt]{article}

\usepackage[margin=1in]{geometry}
\usepackage[T1]{fontenc}
\usepackage{lmodern}
\usepackage{amsmath,amssymb}
\usepackage{graphicx}
\usepackage[font=small,labelfont=bf]{caption}
\usepackage[numbers,sort&compress]{natbib}
\usepackage{bibunits}
\usepackage{etoolbox}
\usepackage[colorlinks=true,linkcolor=blue,citecolor=blue,urlcolor=blue]{hyperref}
\usepackage{xurl}

\newcommand{\hl}[1]{#1}

\newcommand{\E}{\mathbb{E}}
\newcommand{\sinc}{\mathrm{sinc}}

\newcommand{\Tamb}{T_{\mathrm{AMB}}}
\newcommand{\TAMB}{T_{\mathrm{AMB}}}

\newcommand{\K}{\mathrm{k}}

\newcommand{\fb}{R_{\mathrm{b}}}
\newcommand{\J}{\mathrm{j}}
\newcommand{\TFW}{T_{\mathrm{FW}}}
\newcommand{\TBK}{T_{\mathrm{BK}}}
\newcommand{\pcr}{\widetilde{\Delta P}}

\title{First-Principles Channel Model and Capacity-Achieving Transmission Scheme in Thermal Noise Backscatter Links}

\author{Geneva Ecola\textsuperscript{$*$,$\dagger$}, Jida Zhang\textsuperscript{$*$,$\dagger$}, Joseph M. Kahn, and Zerina Kapetanovic\\[1ex]
{\normalsize Department of Electrical Engineering, Stanford University, Stanford, CA, 94305}}

\date{\small Author-accepted manuscript. Published in PNAS 2026, Vol.~123, No.~43, e2606732123.\\
DOI: \href{https://doi.org/10.1073/pnas.2606732123}{10.1073/pnas.2606732123}}

\begin{document}

\maketitle

\renewcommand{\thefootnote}{\fnsymbol{footnote}}
\footnotetext[1]{G.E. contributed equally to this work with J.Z. The order of the co-first authors is alphabetical.}
\footnotetext[2]{To whom correspondence should be addressed. E-mail: gecola@stanford.edu and jidaz@stanford.edu}
\renewcommand{\thefootnote}{\arabic{footnote}}
\setcounter{footnote}{0}

\begin{abstract}
Thermal noise communication systems transmit information by modulating thermal radiation, eliminating the need for a dedicated signal power source and making them attractive solutions for low-power sensor networks. Their efficiency enables long-lived, unobtrusive operation, which is valuable in the Internet of Things, electromagnetic interference-sensitive environments, and RFID tag systems. However, a limited understanding of the wireless channel model and efficient transmission schemes has hindered their practical deployment. Here, we present a wireless channel model that elucidates how systems operating at ambient temperature can modulate thermal noise to convey information. Our results reveal a thermal noise backscatter mechanism that enables significantly higher data rates than previously achieved. We designed and built a thermal noise communication system leveraging this mechanism to achieve data rates in the kilobits-per-second range — over two orders of magnitude higher than prior demonstrations. We also derive the system capacity and show that using polar codes, we are able to achieve this expected capacity. Our channel model and transmission scheme pave the way for new strategies in thermal noise sensing and communication, advancing practical low-power networking for next-generation wireless applications.

\end{abstract}

{\renewcommand{\abstractname}{Significance}
\begin{abstract}
This work develops a wireless channel model and transmission scheme for thermal noise-based communication networks. These systems offer low-power communication in radio-frequency-sensitive environments. By switching a transmitter's impedance, these systems enable wireless communication by inducing changes in the receiver's measured thermal noise temperature. This work models received noise power for impedance-matched and -mismatched transmitter loads. This work uncovers a previously unidentified physical phenomenon of thermal noise backscatter, which explains how noise-temperature changes can arise at the receiver when the entire system is at ambient temperature. Building on this model, this work derives the channel capacity of thermal noise communications and presents a capacity-achieving scheme. The achievable rate is governed by measurement noise bandwidth and thermal contrast between transmitter load states.
\end{abstract}}

\section{Introduction}

Thermal noise is a natural phenomenon that comes in many forms, from Johnson noise in conductors to blackbody radiation. It behaves as white noise throughout the microwave band (up to terahertz frequencies) at temperatures above several kelvin as described by the Rayleigh-Jeans approximation~\cite{Ulaby_1981_MicrowaveRemoteSensing}. Johnson noise arises from the random movement of charge carriers in conductors at non-zero temperatures, delivering tiny amounts of power $P_{\mathrm{n}} = \K TB$ to an impedance matched load where $\K$ is Boltzmann's constant, $T$ is the temperature of the conductor, and $B$ is the noise bandwidth~\cite{Johnson_1928_ThermalAgitation}. Other noise sources, such as cosmic microwave background radiation or noise from active circuits, are characterized by an equivalent noise temperature proportional to the noise power $P_\mathrm{n}$ they deliver. The laws of thermodynamics require that passive devices (e.g., resistors, capacitors, and inductors) have noise temperatures equal to their physical temperature, whereas active circuits can exhibit noise temperatures different from the ambient environment~\cite{VanDerZiel_1954_Noise}. A low-noise amplifier (LNA), for instance, may emit thermal noise from its input that is cooler than its physical temperature~\cite{Frater_1981_ActiveColdLoad, Netzer_1981_LNADesign}.

In electronic systems, engineers often regard thermal noise as a fundamental limitation that constrains a system's signal-to-noise ratio (SNR) or sensitivity. Yet, for decades, thermal noise has enabled sensing applications, including Johnson noise thermometry and radio astronomy. Johnson noise thermometers, systems that measure thermal noise in conductors to infer temperature, and radiometers, radio receivers used to observe radiation from cosmic bodies, employ similar techniques to achieve highly sensitive noise measurements. One strategy for mitigating the impact of random fluctuations in these measurements is switching between the noise source under characterization and a known reference source~\cite{Qu_2019_JohnsonThermometry, Ulaby_1981_MicrowaveRemoteSensing}.

\begin{figure}[!t]
\centering
\includegraphics[width=0.9\linewidth]{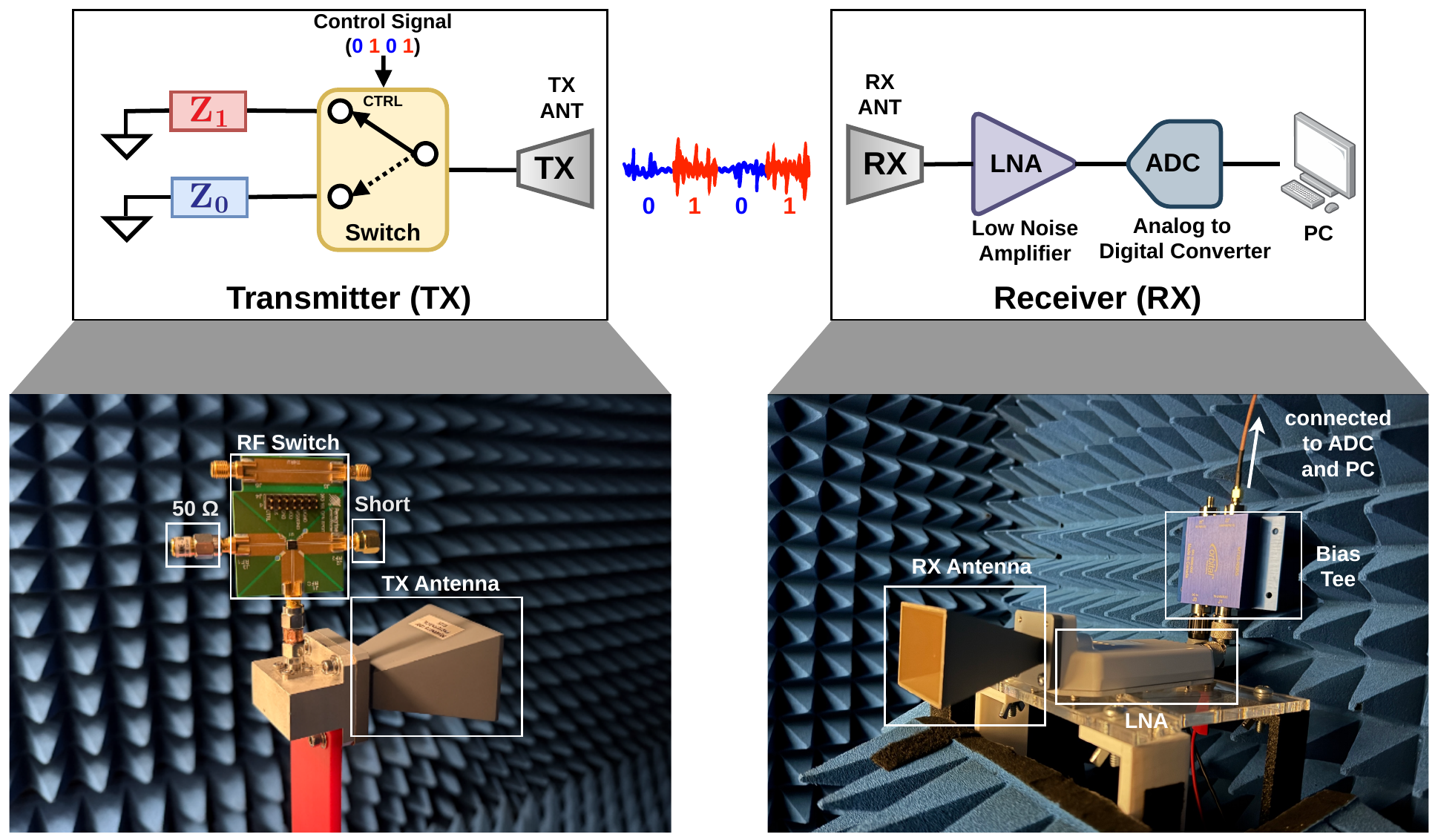}
\caption{\textbf{Overview of a thermal noise communication system.} The diagram shows the transmitter (TX) and receiver (RX) architecture and photographs of the hardware setup in an anechoic chamber; information is encoded by switching between passive loads with different impedance states, which modulates the received thermal noise power.}
\label{fig:SystemOverview}
\end{figure}

The idea of switching between noise sources has inspired research on noise-based communication systems, which treat thermal noise as an always-available carrier for information \cite{Kish_2005_StealthCommunication, Kapetanovic_2022_ModulatedJohnsonNoise}. Fig.~\ref{fig:SystemOverview} illustrates an example of a thermal noise communication system, which consists of a passive transmitter (TX) and a receiver (RX) wherein the TX does not actively transmit radio signals, but instead modulates thermal noise by switching between passive loads with different impedance states. In the example shown in the photograph, the system sends information by switching between a $50~\Omega$ and a short circuit terminator. The nature of the transmitter load alters the thermal noise it radiates, which a highly sensitive radio receiver can wirelessly detect.

Thermal noise communication confers unique advantages on wireless systems due to its unconventional nature. First, it does not require the transmitter or receiver to generate a radio frequency (RF) carrier, thereby making the transmitter ultra-low power and enabling these systems to operate without spectrum licenses. Additionally, unlike ambient backscatter communication, another low-power connectivity solution, thermal noise communication does not need an ambient RF carrier, such as Wi-Fi or TV broadcast signals \cite{Liu_2013_AmbientBackscatter, Kellogg_2014_WifiBackscatter}. As a result, the system is available anywhere, anytime. Finally, due to its low power operation, it produces minimal electromagnetic interference and is therefore less obtrusive than active systems.

This technology is particularly well-suited for applications in which the environment is highly sensitive to RF emissions and where link parameters, such as range, antenna gains, and noise bandwidth, can be adjusted to achieve an SNR that supports the required data rate. Key use cases include wireless communications near radio astronomy observatories, wideband spectrum monitoring deployments, or communication near electromagnetic-sensitive equipment in hospitals. In these settings, even minimal active RF transmissions can interfere with the sensitive signals these systems are designed to detect. By enabling passive communication without active RF emissions, thermal noise communication enables wireless connectivity in previously incompatible environments.

Demonstrations of thermal noise communication show promise, but researchers have not fully characterized how these systems operate ~\cite{Kapetanovic_2022_ModulatedJohnsonNoise, Kapetanovic_2023_CosmicBackscatter, Garman_2024_CarrierFreeRFID}. Many questions remain about the mechanism by which these communication systems, operating entirely at ambient temperature, create thermal contrast in a wireless channel. Existing models assume that switching the transmit antenna's load between two impedance states changes the receiver's measured noise temperature. However, these models do not fully explain the physical origin of this contrast~\cite{Basar_2023_TherMod}. To date, researchers have not proposed a complete wireless channel model that explains this fundamental system phenomenon and is fully validated by experimental results.

In this paper, we address this gap by presenting the first complete wireless channel model for thermal noise communication, revealing fascinating behavior unique to thermal noise channels. We introduce an analytical model of the wireless channel that explains how the transmitter modulates thermal noise using different loads. We divide this model into two load conditions, impedance-matched and -mismatched, because the channel's behavior differs based on this property.
While an impedance-matched load acts as a source of thermal noise that is transferred directly to the receiver, a mismatched load causes the transmitter's antenna to appear as a reflector. This results in cold noise emitted by the receiver's LNA backscattering off the transmit antenna and returning to the receiver. This leads to noise whose temperature deviates from the ambient environment. These observations are used to develop models for the expected thermal noise power measured by the receiver and are supported by numerical and experimental results.

In addition to developing the channel model, we derive the bit-error probability, SNR, and capacity of the thermal noise channel. We demonstrate that capacity is achieved by adopting a high modulation bit rate. While this approach lowers the SNR per symbol, the increased rate of channel usage ultimately maximizes the overall information rate. Finally, we introduce a transmission scheme that achieves the theoretical channel capacity using polar codes, a class of capacity-achieving error correction codes \cite{arikan2009channel}. We implemented a system that achieves the expected capacity to experimentally validate our analyses. The results in this paper represent a fundamental advance over the performance models that researchers have previously assumed and address key barriers to deploying this technology in next-generation sensing, health, and communication networks.

\section{Results}

Thermal noise communication systems send information by modulating the noise power detected by the system's receiver. This noise depends on the impedance and noise properties of the transmit antenna's load. Achieving the highest data rate requires selecting loads that maximize the difference in received power when switching between them. In the following two sections, we derive the expected received thermal noise power for transmitter loads that are impedance-matched and mismatched, so systems can be designed to maximize communication efficiency.

\begin{figure}[!t]
\centering
\includegraphics[width=0.9\linewidth]{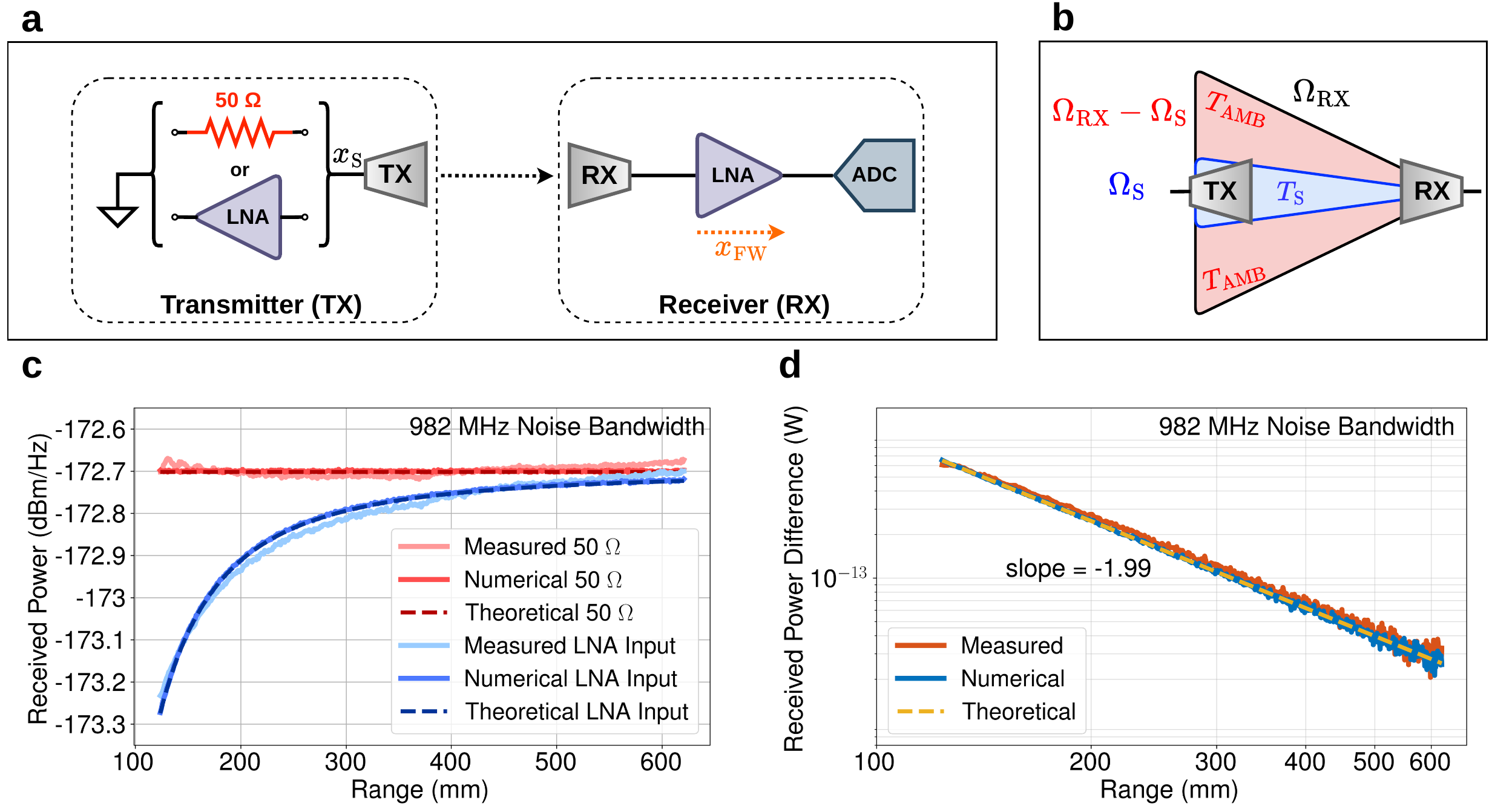}
\caption{\textbf{Wireless channel of thermal noise communication using matched loads. a} The diagram shows the experimental setup used to evaluate the channel behavior of matched loads: a passive $50~\Omega$ and the input of an LNA. The noise from the matched load $x_\mathrm{S}$ travels through the wireless channel to the receiver and gets added to the forward-propagating LNA noise $x_\mathrm{FW}$. \textbf{b} Diagram of the beam angles used to calculate the thermal noise from the source and ambient background sensed by the receiver. \textbf{c} Plots of the theoretical, numerical, and experimental received noise power measured wirelessly with the transmitter connected to the matched loads. \textbf{d} The received power difference between the loads, showing that the backward-propagating noise power from the LNA input decays over the range $r$ as $1/r^{1.99}$.}
\label{fig:MatchedDiagrams}
\end{figure}

\subsection{Channel Model for Matched Load Case}
\label{sec:matched}

When a matched load, i.e., $Z = 50\ \Omega$, is connected to the transmitter, the expected noise power measured by the system's receiver, $P_{\mathrm{MATCH}}$, comprises three independent thermal noise sources: the transmitter's load, thermal background radiation, and receiver noise. The power from the transmitter's load depends on its noise temperature, $T_{\rm S}$. We model the noise voltage at the TX antenna, $x_{\rm S}$, as a zero-mean Gaussian random variable with variance equal to the noise power $\K T_{\rm S} B$. This noise propagates from the TX to the RX through the wireless channel with gain coefficient

\begin{equation}
\begin{aligned}
h=\frac{\sqrt{G_{\mathrm{TX}}G_{\mathrm{RX}}}\lambda \exp(\frac{-\J2\pi r}{\lambda})}{4\pi r},
\end{aligned}
\end{equation}
where $G_{\mathrm{TX}}$ is the transmit antenna gain, $G_{\mathrm{RX}}$ is the receive antenna gain, $\lambda$ is the wavelength of the receiver's center frequency, and $r$ is the distance between the antennas. This expression arises from how wireless signals propagate in free space, with the received power decreasing according to $1/r^2$ as the fixed transmit power is spread uniformly over the surface of a sphere whose area grows as $4\pi r^2$ \cite{friis1946note}. Using this expression, the received noise power is $|h|^2\K T_{\rm S}B$.

While the physical system is entirely at ambient temperature, it is possible to have matched loads with different noise temperatures. Fig.~\ref{fig:MatchedDiagrams}a shows examples of this: a passive 50~$\Omega$ and the input of a powered-on LNA. Analyzing the behavior in both cases enables us to fully understand the wireless channel. The passive 50~$\Omega$ load's noise temperature is equal to ambient, so  $T_{\rm S}=T_{\rm{AMB}}$. By contrast, the powered-on LNA input emits cold thermal noise at temperature $T_{\rm{BK}}$, which we refer to as the \textit{backward-propagating LNA noise}, so $T_{\rm S}=T_{\rm{BK}}$. In a well-designed LNA, $T_{\rm BK}$ is well below ambient temperature; for example, the LNA used in our experiments has $T_{\rm BK}=55$~K.

The RX antenna also receives thermal background radiation at ambient temperature, $T_{\rm AMB}$, from noise in its field of view. Fig.~\ref{fig:MatchedDiagrams}b illustrates the beam solid angle defined by the $-3$~dB points of the receive antenna’s gain pattern, $\Omega_{\mathrm{RX}}$, and the solid angle that the transmit antenna's effective area subtends as seen by the receiver, $\Omega_{\rm S}$. With no transmitter present, the receiver simply receives noise power $kT_{\rm AMB}B$, but with the transmit antenna in the receiver field of view, the background noise then scales by,

\begin{equation}
\begin{aligned}
\frac{\Omega_{\mathrm{RX}}-\Omega_\mathrm{S}}{\Omega_{\mathrm{RX}}}=1-\frac{G_{\mathrm{TX}}G_{\mathrm{RX}}\lambda^2}{(\mathrm{4\pi} r)^2} =1-|h|^2.
\end{aligned}
\end{equation}

In addition to the thermal noise received by the antenna, the RX will also receive noise from its own front-end circuitry. The LNA adds noise as signals propagate from its input to its output, which we refer to as \textit{forward-propagating LNA noise} at noise temperature $\TFW$. Owing to the high gain of the LNA, it dominates the receiver-contributed noise, adding power $\K\TFW B$~\cite{Ulaby_1981_MicrowaveRemoteSensing}.

Adding the three aforementioned noise sources together, the receiver measures a total noise power, in the absence of external interference, given by~\cite{Falconer_2025_WJNT},
\begin{equation} \label{eq:matchedreceivepower}
P_{\mathrm{MATCH}} = \K(T_{\rm S}-T_{\rm AMB})B\tfrac{G_{\mathrm{TX}}G_{\mathrm{RX}}\lambda^2}{(\mathrm{4\pi} r)^2} + \mathrm{k}(\TAMB + \TFW)B.
\end{equation}
\noindent This expression shows that when $T_{\rm S} = \Tamb$, such as when the load is an impedance-matched resistor at room temperature, the expected noise power is independent of antenna gain, wavelength, or range. When $T_{\rm S} \neq T_{\rm AMB}$, the noise power depends on the wireless channel parameters and notably converges to ambient as the distance between the two antennas increases.

To verify this model, we conducted numerical simulations and experimental measurements, as described in the Methods and Appendix S1. The implemented system operates at 11.2~GHz, with an effective noise bandwidth of 982~MHz. We obtained the experimental results using the hardware pictured in Fig.~\ref{fig:SystemOverview} with a $50~\Omega$ terminator and the input of a powered-on LNA as the transmitter's loads.  We simulated random noise voltage samples to generate the numerical results. To align the theoretical and numerical values with the experimental measurements, the system hardware was characterized (Appendix S2). To ensure that measured results were truly thermal noise, control experiments were conducted (Appendix S3).

Both the numerical simulations of the expected noise power and experimental measurements align with the presented theoretical model as the range between the TX and RX is varied. Fig.~\ref{fig:MatchedDiagrams}c shows that when a passive $50~\Omega$ load is connected, the received noise power is constant regardless of range. Whereas when the input of an LNA is connected, the detected noise power is lower, converging to the passive load's power as the range increases. This matches the expected behavior predicted by Eq.~\ref{eq:matchedreceivepower}.

The measured power in both cases fluctuates due to drifts in the hardware's gain and noise figure. Taking the difference between the received power of the two loads at each range position, measured within one second of each other, significantly reduces these fluctuations and enables a precise quantitative comparison between the theoretical model, numerical simulation, and experimental results. Fig.~\ref{fig:MatchedDiagrams}d plots the difference between the powers received using the two loads. The power difference scales as $1/r^{1.99}$, in strong agreement with the theoretical model, which predicts $1/r^2$. On average, the theoretical and experimental power ratios differ by only 0.01~dB, demonstrating that the proposed wireless channel model accurately describes the system.

This section demonstrates how matched loads can cause measurable thermal contrast at the receiver based on differences in their noise temperatures. In the two loads from the experiment, this difference was enabled by the power supplied to the LNA, which enabled it to have a noise temperature that differed from the ambient temperature. As a result, the the two loads could be used as a thermal noise communication link. Yet, enabling an ultra-low power transmitter requires that only passive loads are used. But, is it possible to measure noise temperatures that differ from the ambient environment using only passive loads? In the next section, we demonstrate how changing the transmitter load impedance enables this.

\subsection{Channel Model for Mismatched Load Case}

\begin{figure}[!t]
\centering
\includegraphics[width=0.9\linewidth]{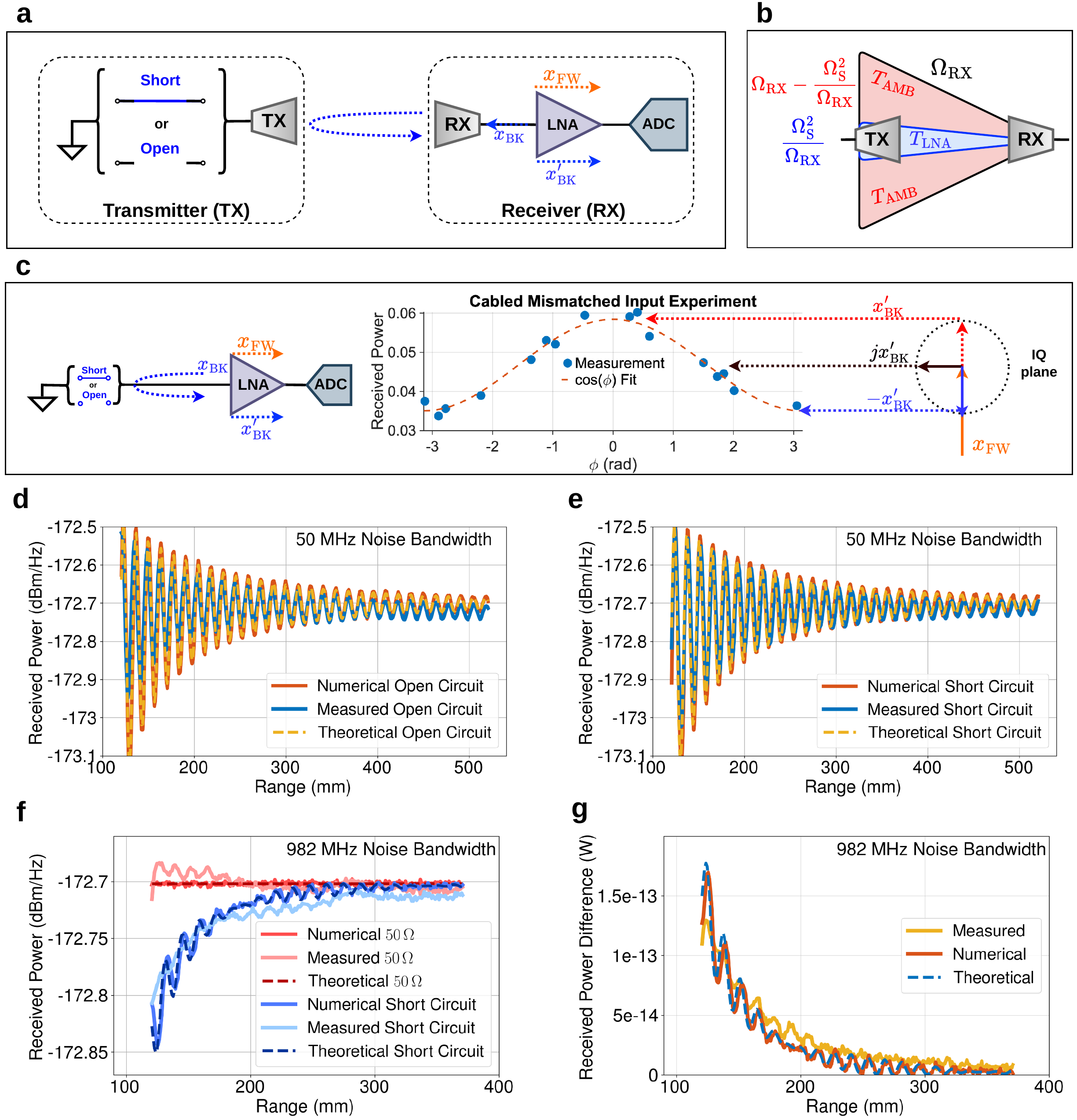}
\caption{\textbf{Wireless channel of thermal noise communication using mismatched loads. a} The diagram shows the experimental setup used to evaluate the channel behavior of mismatched loads: open and short-circuit terminators. The mismatched loads cause $x_\mathrm{BK}$ to reflect back to the receiver as $x'_\mathrm{BK}$, where it combines with $x_\mathrm{FW}$. \textbf{b} Diagram of the beam angles used to calculate the thermal noise sensed by the receiver. \textbf{c} Cabled experiment and IQ plane illustration showing the combination
of $x'_\mathrm{BK}$ and $x_\mathrm{FW}$. The units of received power are arbitrary. \textbf{d} The received power, as range is varied, of the transmitter with an open circuit load using a 50~MHz bandwidth. \textbf{e} The received power, as range is varied, of a transmitter with a short circuit load using a 50~MHz bandwidth. \textbf{f} The received power, as range is varied, of the transmitter with $50~\Omega$ and short circuit loads using a 982~MHz bandwidth. Here, the 50 $\Omega$ terminator acts as a reference noise source at room temperature. \textbf{g} The difference in received power, as range is varied, between the transmitter terminated in a $50~\Omega$ load and a short circuit using a 982~MHz bandwidth.}
\label{fig:MismatchedDiagrams}
\end{figure}

When the transmitter's load is mismatched, such as a short ($Z=0$) or open  ($Z=\infty$) circuit with reflection coefficient $|\Gamma| = 1$, the received thermal noise power, $P_{\rm MIS}$, differs significantly from the matched case. This occurs because the mismatched load causes the transmit antenna to act as a reflector of incident thermal radiation, as depicted in Fig.~\ref{fig:MismatchedDiagrams}a and b, rather than another source. This causes thermal radiation within the TX antenna's field of view to be backscattered to the receiver, including ambient background radiation and even backward-propagating LNA noise emitted by the input of the receiver's LNA. We split our analysis of the received thermal noise power into two parts: first, the noise contributions from the LNA, including both forward- and backward-propagating noise, and second, the contribution from ambient background radiation.

First, we analyze the receive LNA's contributions. Consider at time $t$, the receiver's backward-propagating LNA noise, with voltage $x_{\rm BK}(t)$,  backscatters off the transmit antenna and returns to the receiver as $x'_{\rm BK}(t)=h^2\exp(\J\phi)x_{\rm BK}(t-\tau)$. Here, $\phi = \angle \Gamma$ is the phase angle of the transmitter's reflection coefficient, and $\tau = 2r/c$ gives the two-way propagation delay for range $r$, where $c$ is the speed of light. When the reflected signal arrives at the receiver, it combines with forward-propagating LNA noise modeled as an input-referred voltage source, $x_{\rm FW}(t)$. This has two significant consequences:
1) Since $T_{\rm BK} \neq T_{\rm AMB}$, it creates thermal contrast within the wireless channel as compared to a passive matched load.
2) Since the same LNA circuitry generates $x_{\rm BK}$ and $x_{\rm FW}$, there is high correlation between $x_{\rm FW}$ and $x'_{\rm BK}$, causing the signals to constructively and destructively interfere with one another.
Together, these effects enable the system to measure noise temperatures that differ from the ambient environment.

To calculate the power from both LNA noise sources, we sum their voltages and then convert this sum to power, as shown in Eq.~\ref{eq:P_LNA}. This yields three distinct contributions: (1) the LNA's forward-propagating noise, (2) the LNA's backscattered, backward-propagating noise, and (3) the cross-correlation term representing the interference between these two signals. A detailed derivation is included in Appendix S4.
\begin{equation}
\label{eq:P_LNA}
\begin{aligned}
P_{\rm LNA}
&= \E[|x_{\rm FW}+x'_{\rm BK}|^2] \\
&= \K \TFW B
 + \K \TBK B\tfrac{G_{\mathrm{TX}}^2G_{\mathrm{RX}}^2\lambda^4}{(4\pi r)^4} + 2\K \sqrt{\TFW \TBK}B\cos\!\left(\psi+\phi-\tfrac{4\pi r}{\lambda}\right)
\sinc(B\tau)\tfrac{G_{\mathrm{TX}}G_{\mathrm{RX}}\lambda^2}{(4\pi r)^2}.
\end{aligned}
\end{equation}
In Eq.~{\ref{eq:P_LNA}}, $\psi$ is the phase of the cross-correlation between $x_{\rm FW}$ and $x_{\rm BK}$, determined by the internal LNA circuitry. The non-zero cross-correlation term causes oscillations in the received noise power as the range or center frequency is varied. The magnitude and phase of this term depend on the measurement noise bandwidth of the receiver and the range between the RX and TX, as they govern the degree of constructive and destructive interference between the two band-limited Gaussian processes.

A simple cabled experiment provides intuition for this phenomenon: we connect the LNA input to a cable terminated in either an open or a short circuit (setup shown in Fig.~\ref{fig:MismatchedDiagrams}c) and measure the power received. By varying the length of the transmission line, we alter the phase shift, $\phi$, between the backscattered, backward-propagating LNA noise and the forward-propagating LNA noise. As in Eq.~\ref{eq:P_LNA}, Fig.~\ref{fig:MismatchedDiagrams}c demonstrates how the sum of these two correlated noise signals can combine constructively or destructively, depending on their phase relationship. This is exactly the mechanism by which the wireless channel exhibits power oscillations as the transmitter range is varied, a process directly analogous to altering the cable length.

In a wireless system, $P_{\mathrm{LNA}}$ accounts for one part of the total noise power detected by the receiver. However, as in the matched load case, the receiver also detects thermal noise from the ambient environment. In the mismatched case, it arises from both the background in view of the receiver and ambient thermal noise backscattered by the transmit antenna, as shown in Fig.~\ref{fig:MismatchedDiagrams}b. This results in additional noise
$P_{\rm AMB} =\K \TAMB B(1-|h|^4)$.

Since the LNA and ambient noise are not correlated, the total received power, $P_{\rm MIS}$, from these two sources can be added together, as shown in Eq.~\ref{eq:P_MIS}.
\begin{equation}
\label{eq:P_MIS}
\begin{aligned}
P_{\rm MIS}&= \K(\TAMB + \TFW)B + \K(T_{\rm BK}-T_{\rm AMB})B\tfrac{G_{\mathrm{TX}}^2G_{\mathrm{RX}}^2\lambda^4}{(4\pi r)^4} \\
&\quad + 2\K\sqrt{\TFW T_{\rm BK}}B\cos\!\left(\psi+\phi-\tfrac{4\pi r}{\lambda}\right)
\sinc(B\tau)\tfrac{G_{\mathrm{TX}}G_{\mathrm{RX}}\lambda^2}{(4\pi r)^2}.
\end{aligned}
\end{equation}
Depending on the system's conditions, different terms will dominate the received power. At very large ranges, the receive noise power tends to $\K(\TAMB + \TFW)B$, the same as in the matched case from Eq.~\ref{eq:matchedreceivepower}.
At close range, the behavior depends on the ratio between the sampling time, $1/B$, and the backscattered signal's time delay, $\tau$.

\paragraph{Low Noise Bandwidth:} When $1/B > 3\tau$, then $\sinc(B\tau) > 0.8$. This causes the final oscillating term to dominate the expected behavior since its decay is $1/r^2$, while the $\TBK-\TAMB$ term falls off with $1/r^4$. Thus, the expression is dominated by the final oscillating term, centered at $\K (\TAMB + \TFW)B$.

We performed numerical and experimental evaluations with open and short circuit terminators using the hardware shown in Fig.~\ref{fig:SystemOverview}. The system used a measurement noise bandwidth of 50~MHz corresponding to a sampling period of 20~ns. We varied the RX-TX range, resulting in $\tau = 3.5$ to $5.5$~ns, with an additional 2.8~ns delay from electrical lengths. With this setup, the sampling period was three to four times longer than the propagation delay. Fig.~\ref{fig:MismatchedDiagrams}d and Fig.~\ref{fig:MismatchedDiagrams}e show the dominant oscillating behavior with the open and short circuit loads. The average error between the theoretical and measured results is 0.02~dB, closely matching the oscillation periods and $1/r^2$ envelope. The nulls of the open and short measurements are offset by $90^\circ$, which is the measured phase difference between the reflection coefficients of the two loads at the system's center frequency. While on a much smaller scale, this mechanism also causes slight power oscillations when measuring the transmitter with a matched load due to the structural mode backscatter from the TX antenna; we describe this in Appendix S5.

\paragraph{High Noise Bandwidth:} When $1/B < \tau/3$, then $\sinc(B\tau) < 0.1$. With the sampling rate sufficiently high, interference from the backscattered, backward-propagating LNA noise vanishes. When this happens, Eq.~\ref{eq:P_MIS} reduces to the first two terms, and the received power converges to ambient according to $1/r^4$. This can be understood as using a sufficiently wide bandwidth so that the peaks and nulls of the different frequencies within the bandwidth cancel, yielding a smoother received power as the range varies.

We conducted an experiment with a noise bandwidth of 982~MHz, corresponding to a 1~ns sampling period. As shown in Fig.~\ref{fig:MismatchedDiagrams}f, when the noise bandwidth is increased, the oscillatory behavior in the noise power for the open circuit disappears. Instead, the received noise power converges with the ambient $50~\Omega$ as the range is increased. The difference between the power received from a $50~\Omega$ and the open circuit is plotted in Fig.~\ref{fig:MismatchedDiagrams}g approximately following a $1/r^4$ trend, with an average theory prediction error of 0.01~dB.

Summarizing the above, we have presented an operating mechanism that explains how the receiver can measure noise temperatures that differ from the ambient environment using only passive loads. The key enabler is the cold thermal noise appearing at the receiver's LNA input. This allows us to modulate the receiver's thermal noise using a passive transmitter.

\subsection{Channel Performance}

\begin{figure}[!t]
\centering
\includegraphics[width=0.9\linewidth]{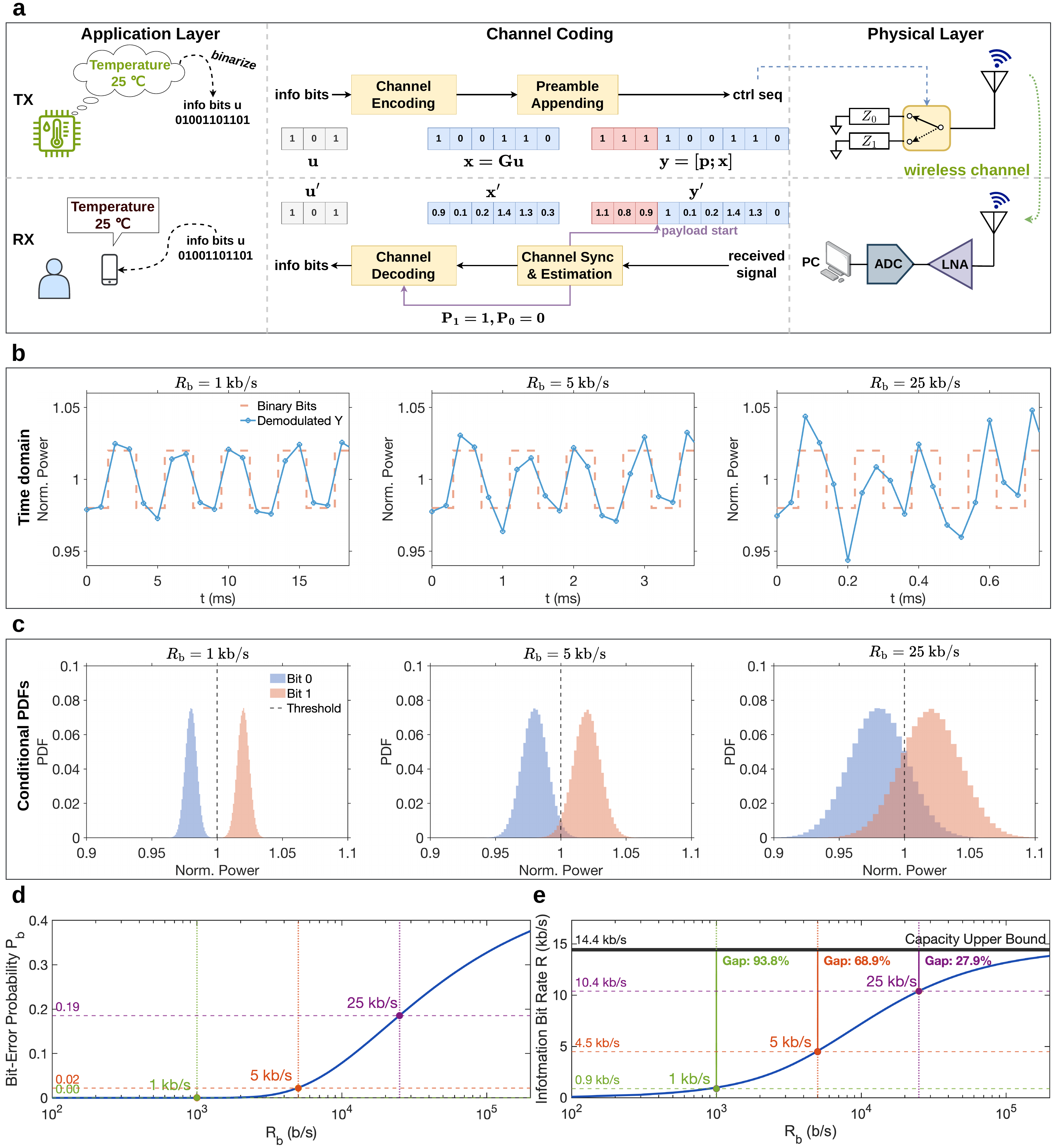}
\caption{\textbf{Proposed thermal noise communication system. a} End-to-end system diagram. After acquiring sensor data, the sensor converts it into bits and encodes it as a binary control signal that drives an ultra-low-power switch. The end user then observes corresponding changes in the received noise power and decodes the transmitted message accordingly. \textbf{b} As the modulation bit rate $\fb$ increases, the bit interval shortens and fewer noise samples are averaged per bit, increasing the effective noise variance. \textbf{c} The conditional probability densities for bits 0 and 1 overlap more, yielding higher uncoded bit-error probability. \textbf{d} The bit-error probability increases as a result of the reduced integration time per bit. \textbf{e} In spite of this, the information bit rate $R$ also increases as modulation bit rate $R_{\rm b}$ increases. With error correction, the coding gap narrows and the achieved throughput approaches capacity.}
\label{fig:capacity_graph}
\end{figure}

In this section, we analyze the maximum achievable throughput of thermal noise communication systems and describe the implementation of a capacity-achieving link. This system represents an advance over prior research by exploiting the backscatter mechanism identified in this work to increase the system's achievable capacity. Two phase-offset mismatched loads were used, which maximize the received thermal contrast. The system employs a transmission scheme using polar codes that realizes the link capacity and is efficient for use in any thermal noise communication system.

Fig.~\ref{fig:capacity_graph}a shows a full system from an example application layer to the channel coding to the physical layer. The transmitter acquires raw sensor data, binarizes it, encodes the bits using a channel code, and prepends a preamble to form a data packet. The system transmits the packet by toggling the switch between the loads with impedances $Z_0$ and $Z_1$, thereby modulating the thermal noise power between two binary states, which we denote as $P_0$ and $P_1$. The receiver resolves the state by sampling the thermal noise with bandwidth $B$ and averaging the measured noise power over each bit interval. Packet detection, channel estimation, and channel decoding then recover the information bits and ultimately reconstruct the original sensor data.

The system's performance is determined by the following three factors:
\begin{enumerate}
    \item[(1)] The power contrast ratio between two states 0 and 1, defined as $\pcr = (P_1-P_0) / (P_1+P_0)$. We assume $P_1>P_0$ without loss of generality, such that $\pcr>0$. A higher $\pcr$ enhances the detectability and reliability of the transmitted packets at the receiver; the physical setup inherently determines this ratio.
    \item[(2)] The modulation bit rate $R_{\mathrm{b}}$, which can be controlled and optimized at the transmitter to maximize throughput. While increasing $\fb$ facilitates faster transmissions, the shorter integration interval leads to a noisier power observation $Y$ for detecting the transmitted bit $X$ (Fig.~\ref{fig:capacity_graph}b). We show below that the channel capacity increases with $\fb$.
    \item[(3)] The noise bandwidth $B$, which determines the overall noise power that the receiver is able to capture and the receiver's sampling rate. For a given power contrast ratio, larger $B$ improves detectability of the transmitted packet and increases the achievable link capacity.
\end{enumerate}

Next, we present the key performance metrics of the thermal noise channel—including the bit-error probability, capacity, and SNR—all of which depend on $\pcr$ and $\fb$ for a given noise bandwidth $B$. In this paper, we model the noise as a band-limited white Gaussian process with constant power spectral density (PSD) across the bandwidth $B$. A detailed derivation can be found in Appendix S6. For $B \gg R_\mathrm{b}$, the conditional distribution of the averaged power converges to a Gaussian with probability density function (PDF): $p_{Y|X}(y|i) = {\cal N}(y; P_i, \frac{\fb}{B}P_i^2)$, for $i\in \{0,1\}$. The variance is proportional to $\fb$ as illustrated in Fig.~\ref{fig:capacity_graph}c. Assuming equiprobable source bits $\Pr(X=0)=\Pr(X=1)=1/2$ and if $P_0$ and $ P_1$ are at least approximately equal, the bit-error probability is~\cite{Basar_2023_TherMod}
\begin{equation}
\label{eq:Pb}
P_{\rm b} = (1/2)\, \mathrm{erfc}(\sqrt{B / (2R_\mathrm{b})}\cdot \pcr).
\end{equation}
This expression reveals how $\pcr$, $R_\mathrm{b}$ and $B$ determine the system's ability to distinguish between binary states: a higher contrast ratio $\pcr$ or noise bandwidth $B$ improves detection, while a higher modulation bit rate $R_\mathrm{b}$ increases noise and degrades performance. The relationship between the modulation bit rate and bit-error probability is plotted in Fig.~\ref{fig:capacity_graph}d.

Assuming the source bits are equiprobable, the capacity per channel use can be expressed as $C =  H(Y) - H(Y|X)$, where $H$ is entropy. Here, $H(Y)=\int_{-\infty}^\infty p_Y(y)\log_2 \frac{1}{p_Y(y)}{\rm d}y$ with $p_Y(y)=\frac{1}{2}p_{Y|X}(y|0)+\frac{1}{2}p_{Y|X}(y|1)$ and the conditional entropy is the average of the two states: $H(Y|X) = \frac{1}{2}H(Y|X=0)+\frac{1}{2}H(Y|X=1)$. The maximum achievable information rate $R$ is the product of the channel uses per second (the modulation rate $\fb$) and the capacity per channel use ($C$):
\begin{equation}
\label{eq:R}
\begin{split}
R
&= \fb C \\
&= \fb \int_{-\infty}^{\infty} p_Y(y)\log_2\!\tfrac{1}{p_Y(y)}\,\mathrm{d}y \\
&\hspace{1.5em}- \fb\left[
  \tfrac{1}{2\ln 2}
  +\tfrac{1}{4}\log_2\!\left(\tfrac{2\pi \fb P_0^2}{B}\right)
  +\tfrac{1}{4}\log_2\!\left(\tfrac{2\pi \fb P_1^2}{B}\right)
\right].
\end{split}
\end{equation}

While the integral in Eq.~\ref{eq:R} does not admit a closed-form solution, Fig.~\ref{fig:capacity_graph}e plots numerical results showing that as the modulation rate increases, the information rate $R$ increases monotonically with $\fb$. This indicates that the upper bound of $R$ is reached in the low-SNR and high-bit-error probability regime as $\fb\rightarrow \infty$.

In a wireless thermal backscatter link, $P_0$ and $P_1$ are nearly equal, so the conditional PDFs for bits 0 and 1 have nearly equal variances, and the channel is well approximated by an additive white Gaussian noise (AWGN) channel with binary inputs. When the SNR per symbol is sufficiently low, the binary-input AWGN channel capacity is well-approximated by the capacity of a continuous-input AWGN channel with the same SNR per symbol. The latter channel has a capacity per use $C_{\rm AWGN} = (1/2)\log_2(1+{\rm SNR})$, where ${\rm SNR}=(\pcr)^2 B / \fb$ \cite{shannon1948mathematical}. This approximation is highly accurate for $C_{\rm AWGN} < 0.5$. Thus, an upper bound on the information rate of the thermal backscatter link, $\overline{R}$, can be expressed as
\begin{equation}
\label{eq:capacity}
\begin{aligned}
\overline{R}
&= \lim_{\fb \rightarrow \infty} \fb C \\
&= \lim_{\fb \rightarrow \infty} \fb C_{\rm AWGN} \\
&= \lim_{\fb \rightarrow \infty} \tfrac{1}{2}\fb\log_2\!\left(1+\tfrac{(\pcr)^2 B}{\fb} \right) = \tfrac{(\pcr)^2 B}{2\ln 2}.
\end{aligned}
\end{equation}
The rate $\overline{R}$ is achievable via binary modulation, which is capacity-achieving in the low-SNR regime. Note that the information rate upper bound $\overline{R}$ is proportional to the
noise bandwidth, $B$, times the square of the power contrast ratio, $\pcr$. We observe the following:

\begin{itemize}
    \item At low $\fb$ (Fig.~\ref{fig:capacity_graph}b-c left graph), longer bit durations ensure high SNR (which is inversely proportional to $\fb$) and therefore low bit-error probability. However, since $\fb$ is well below the information bit rate upper bound, the system is \textit{modulation bit rate-limited}, not SNR-limited. Since the information bit rate $R$ cannot exceed the modulation bit rate $\fb$, this results in a significant gap to the capacity upper bound $\overline R$. Although this gap could be reduced using non-binary modulation, the upper bound $\overline{R}$ can only be achieved in the high-$\fb$ regime.
    \item At high $\fb$ (Fig.~\ref{fig:capacity_graph}b-c right graph), each bit has a low SNR and therefore high bit-error probability. This is when the system is \textit{SNR-limited}, but not modulation bit rate-limited. The information bit rate is increased because the system uses the channel more frequently and effectively corrects errors caused by the low SNR using capacity-achieving error-correcting codes. Here, the system can achieve the maximum capacity per channel usage using binary modulation.
\end{itemize}

\begin{figure}[!t]
\centering
\includegraphics[width=0.9\linewidth]{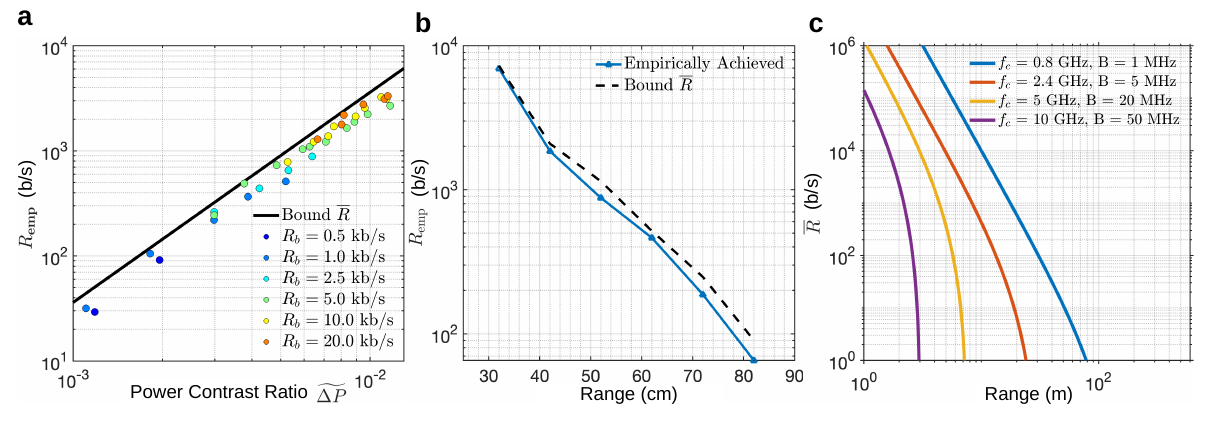}
\caption{\textbf{Performance of a thermal noise communication system. a} Empirically achieved information bit rate $R_{\rm emp}$ measurements with different power contrast ratios $\pcr$. \textbf{b} Empirically achieved information bit rate $R_{\rm emp}$ at different ranges after optimizing center frequency. \textbf{c} Simulated information rate upper bound $\overline{R}$ versus range results for different center frequencies $f_{\rm c}$ and bandwidth $B$ with 20~dBi TX and RX antenna gains.}
\label{fig:capacity_fig}
\end{figure}

We implemented a thermal noise communication system as described above (see Appendix S7 for details) using a noise bandwidth of $B=50~{\rm MHz}$ (which is limited by the sampling rate of the software-defined radio), for a range of power contrast ratios $\pcr\in[10^{-3},10^{-2}]$. In the experimental setup, the measured noise PSD is flat across the channel bandwidth, consistent with the white noise assumption in the analysis above. Owing to the cosine term in Eq.~\ref{eq:P_MIS}, the thermal noise power contrast is maximized when the two loads are both mismatched with relative phase offset $\phi = \pi$. Therefore, open and short circuits are chosen as the two transmitter loads. Since the signal power in Eq.~\ref{eq:P_MIS} also oscillates with frequency and the thermal noise spectrum is inherently ultra-wideband, the receiver scans the entire 982~MHz LNA bandwidth and selects the 50~MHz band that maximizes $\pcr$, and thus capacity. We use polar codes as the error-correcting code due to their simple encoding algorithm and capacity-achieving property. We evaluate the empirically achieved information bit rate, $R_{\rm emp}$, under reliable (error-free) communication conditions. The results demonstrate that the system maintains a small average performance gap $\Delta~=~-0.76~\textrm{dB}$ relative to the maximum information bit rate $R$ for different modulation bit rates $R_{\rm b}$ and power contrast ratios $\pcr$ (Fig.~\ref{fig:capacity_fig}a). Here, $R$ is well-approximated by the low-SNR capacity formula as $R \approx \frac{1}{2}R_{\rm b}\log_2(1+{\rm SNR}) $ and $\Delta \in (0,1)$ is defined as an SNR degradation factor such that the empirically achieved information bit rate is expressed as
$ R_{\rm emp} = \frac{1}{2}R_{\rm b}\log_2(1+\Delta\cdot {\rm SNR})$.

We evaluate throughput as a function of range, as shown in Fig.~\ref{fig:capacity_fig}b. The system sustains $\sim$1~kb/s at 50~cm and 65~b/s at 80~cm. To evaluate the communication range for different center frequencies, we simulated a scenario with two mismatched loads at the transmitter, with a relative phase difference of $\phi = \pi$. By applying Eq.~\ref{eq:P_MIS} and Eq.~\ref{eq:capacity}, Fig.~\ref{fig:capacity_fig}c shows that a communication range of 80~m is achievable at 800~MHz center frequency while maintaining a throughput of 1~b/s.

\section{Discussion}
This paper presents a first-principles channel model for wireless thermal noise communication. The model shows that the channel's behavior is governed by the impedances and noise temperatures of the loads connected to the transmitter's antenna, as well as the noise temperature of the receiver's LNA. Notably, we explain how a thermal noise communication system using a fully passive transmitter operating at ambient temperature can create thermal contrast and transmit information. The key is the cold backward-propagating noise emitted by the receiver's LNA. It backscatters off the TX antenna when the transmitter's load is mismatched. The backscattered, backward-propagating noise is correlated with the forward-propagating LNA noise, creating constructive and destructive interference in the received power.

Several kinds of thermal noise communication are possible depending on the loads the transmitter uses. Prior work has often utilized switching between passive matched and mismatched loads. In this work, we explore a new type of thermal noise backscatter communication architecture that uses only mismatched loads. Using two mismatched loads with offset phases results in a large thermal contrast, owing to the superposition of correlated LNA noise at the receiver. This enables significantly greater differences in received thermal noise power than switching between a passive matched and mismatched load.

In this work, we also derive the system's information-theoretic capacity, which can be used to determine the maximum achievable data rate for any two arbitrary transmit loads. The system implemented in this paper, using two mismatched loads, approached the theoretically calculated capacity and achieved data rates two orders of magnitude higher than prior work. In this paper, demonstrating completeness and correctness of the models and maximizing data rate were prioritized over application-specific practicality. The high-gain antennas and high center frequency enabled detailed measurement of the channel over short ranges. Experiments were conducted in an anechoic chamber, which eliminated interference from outside signals. Future work should focus on making thermal noise communication practical for real deployments.

One major area of work is ambient interference mitigation in channels that are not necessarily thermal-noise-limited. While radio-astronomy bands, which prohibit active transmissions and are therefore interference-free, could in principle be exploited, most wireless spectrum is actively used and subject to frequent interference. Because the passive matched and mismatched loads are inherently broadband, and thermal noise is white at radio frequencies, the TX modulates thermal noise across the RX's entire bandwidth. Interference can therefore be mitigated by sampling the receiver's full noise bandwidth and selectively filtering out interference that partially occupies the spectrum.

Depending on system requirements, including range and size, different center frequencies and antenna sizes can be weighed against achievable throughput for the application. Achieving higher throughput over longer ranges requires greater antenna gain, which generally requires antennas with larger effective apertures and more accurate pointing. Because the received thermal contrast varies with both range and frequency, changes in link distance require re-optimizing the center frequency to maintain the maximum achievable information rate. For very low-bandwidth sensor data or messaging, it may be acceptable to reduce the achievable capacity in favor of a smaller form factor. These tradeoffs make the technology particularly suitable for low-data-rate sensing and messaging in radio-frequency-sensitive environments where low transmitter power and minimal electromagnetic interference are priorities, such as hospitals or near spectrum monitoring devices like radio astronomy observatories. Beyond data transfer, this architecture also supports sensing: range-based localization can be achieved by estimating the characteristic spacing of oscillation patterns. Overall, this work supports a growing body of research on how noise-based communication systems can enable new and exciting ways to communicate for next-generation wireless applications.

\section{Materials and Methods}
\label{sec:methods}

\subsection{Experimental Setup}
We conducted experiments in an anechoic chamber built on an optical breadboard table using metal sheets and RF absorber foam. We 3D-printed antenna mounts for the transmitter and receiver to precisely align the setup and covered it in absorber foam to reduce reflections. To perform range experiments, we used an automated stage programmed to move the transmit antenna with 0.1~mm accuracy. All experiments were conducted at the lab's ambient temperature, which ranged from 290~K to 300~K.

\subsection{Hardware}
Fig.~\ref{fig:SystemOverview} shows the system's hardware. The transmitter used a 15~dBi horn antenna, an RF switch, and two loads, which were modified for each experiment, including a $50~\Omega$ terminator, a low-noise block (LNB), an open circuit terminator, and a short circuit terminator. The receiver was implemented using the same horn antenna and an LNB that simultaneously amplifies and down-converts the RF signal to an intermediate frequency for sampling by a software-defined radio or analog-to-digital converter.

\subsection{Signal Processing}
The raw complex ADC samples were converted to power and integrated to evaluate the wireless channel model and channel capacity. For wireless channel characterization, we integrated 100 million samples for each point. The average noise power measured with a matched $50~\Omega$ terminator was used for calibration, i.e., converting noise temperature to power in ADC arbitrary units.

\subsection{Communication System Implementation}
In the capacity experiments, each packet contains a 2047-bit preamble implemented as an $m$-sequence (maximum-length sequence) and a 4096-bit payload encoded with polar codes~\cite{arikan2009channel}. For each power contrast ratio $\pcr$, we sweep multiple bit rates $\fb$ and select code rates based on capacity estimates for the corresponding channel. The information bits are i.i.d. Bernoulli-$1/2$. At the receiver, we correlate the received power sequence with the preamble to detect the packet start and jointly estimate the power of each state. Using these estimates, we compute log-likelihood ratios for the payload bits and decode via CRC-aided successive-cancellation list decoding~\cite{tal2015list}. The bit-error probability is obtained by comparing the decoded payload with the transmitted information bits.

\subsection{Numerical Simulations}
Numerical simulations were conducted by generating random noise voltages from a complex Gaussian distribution for each noise source based on its expected noise temperature. The matched and mismatched cases used the same noise temperatures for the ambient background and receiver noise, while the noise from the transmitter was modified based on impedance state. These samples were then processed using the expected hardware specifications, system bandwidth, and integration time.

\section*{Data, Materials, and Software Availability}
The data and code supporting the findings of this study are publicly available on Github at \url{https://github.com/S4-Lab/first-principles-thermal-noise-backscatter.git}.

\section*{Author Contributions}
All authors conceived of the work and contributed to the theoretical analysis. J.Z. and G.E. performed the experiments and collected the data. All authors contributed to the writing and revision of the manuscript.

\section*{Competing Interests}
The authors declare no competing interests.

\section*{Acknowledgments}
The authors appreciate technical discussions and manuscript feedback from Professor Joshua Smith and Dr. Shanti Garman.

\bibliographystyle{unsrt}
\bibliography{refs}

\makeatletter
\def\@extra@binfo{@SI}
\def\@extra@b@citeb{@SI}
\makeatother
\begin{bibunit}[unsrt]
\apptocmd{\appendix}{\section*{Supplementary Information Appendix}}{}{}
\clearpage
\appendix
\renewcommand{\thesection}{S\arabic{section}} 
\renewcommand{\theequation}{S\arabic{equation}} 
\renewcommand{\thefigure}{S\arabic{figure}}     
\renewcommand{\thetable}{S\arabic{table}}       
\numberwithin{equation}{section}                
\setcounter{section}{0}
\setcounter{equation}{0}
\setcounter{figure}{0}
\setcounter{table}{0}
\setcounter{section}{0}

\section{Experimental Hardware}
\label{suppsec:experiment}
Both the transmitter \hl{(TX)} and receiver \hl{(RX)} use 15~dBi gain, WR-75 waveguide horn antennas (Pasternack PEWAN075-15) \cite{pasternack_wr75_horn}. To meet varying experimental requirements, two different TX switches were employed: a wideband absorptive mechanical \hl{radio frequency} (RF) switch (Mini-Circuits MSP2T-18XL+) \cite{minicircuits_msp2t_18xl} was selected for its low insertion loss and high isolation in the experiments shown in Fig. 2c–d and Fig. 3d–g; while an electronic RF switch (pSemi EK42520-03) was used for the results in Fig. 5a–b due to its superior switching speed \cite{psemi_ek42520_03}. The receiver front end incorporates a low-noise block (LNB) down converter (Orbital Research 5300 Off-the-Shelf Ku-band PLL LNB) \cite{orbital_5300_lnb}. The LNB is powered using a bias tee (Orbital Research Mux Tee)\cite{orbital_biastee}. Signal digitization was performed using either a USRP B210 software-defined radio (SDR) \cite{usrp_b210} (Fig. 3d–e and Fig. 5a–b) \hl{for the 50~MHz bandwidth experiments} or a TI ADC12J4000EVM 12-bit RF-sampling \hl{analog-to-digital converter} (ADC) \cite{ti_adc12j4000evm} (Fig. 2c–d and Fig. 3f–g) \hl{for the 982~MHz bandwidth experiments}. The sampling rate of the TI ADC was set to 2.7 GS/s. When the ADC was used, additional gain was required, so an additional broadband \hl{low-noise amplifier} (LNA) (Mini-Circuits ZX60-P105LN+) was added at the output of the bias tee \cite{minicircuits_ZX60-P105LN+}.

\section{System Characterization}
\label{suppsec:systemcharacterization}
In this section, we describe the system characterization process to obtain the system parameters including forward- and backward-propagating LNA noise temperatures $T_{\rm BK}$, $T_{\rm FW}$, the antenna gains $G_{\rm TX}$, $G_{\rm RX}$, the antenna phase center offset used to determine range $r$, and the measurement noise bandwidth $B$.

\subsection{LNA Noise Temperature Characterization}
\begin{figure}[!h]
    \centering
    \includegraphics[width=0.7\linewidth]{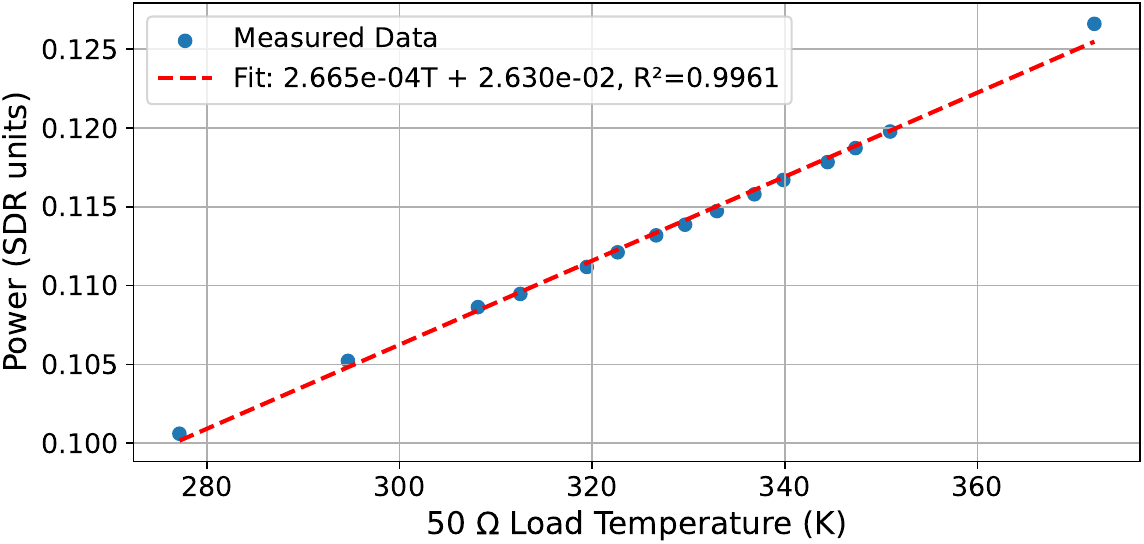}
    \caption{LNA noise temperature characterization experiment.}
    \label{suppfig:P_vs_T}
\end{figure}

We utilize the USRP SDR to characterize the forward- and backward-propagating LNA noise temperatures. To perform this characterization, \hl{two measurement setups were used. First,} a $50~\Omega$ terminator is connected to the LNA input, and \hl{its output is connected to the SDR.} The received power \hl{was} measured as we \hl{varied the terminator's physical temperature}. The power measured by the SDR, $P_{\mathrm{SDR}}$, is modeled as a linear function of the load temperature $T_L$:
\begin{equation}
    P_{\mathrm{SDR}} = G_{\mathrm{SDR}}(kT_LB + kT_{\mathrm{FW}}B) = aT + b
\end{equation}
where $G_{\mathrm{SDR}}$ is the system gain, \hl{including the LNA,} that converts analog power to SDR-defined digital units, $k$ is the Boltzmann constant, $B$ is the measurement bandwidth, and $a$ and $b$ are the slope and intercept of the linear fit, respectively. \hl{Using the data from} \autoref{suppfig:P_vs_T}, \hl{a and b were estimated. Based on these measurements $T_{\rm FW} = b/a = 99~\rm K$.}

\hl{Next, a second LNA was placed between the $50~\Omega$ terminator and the first LNA, with its output connected to the terminator and its input connected to the first LNA's input. This allows us to measure the noise temperature of the LNA's input.  Using this setup, we measured} $P_{\rm BK} = 0.041$, from which we know that $T_{\rm BK} = (P_{\rm BK} - b) / a = 55~\rm K$.

\subsection{Antenna Characterization}
\begin{figure}[htbp]
    \centering
    \includegraphics[width=0.9\linewidth]{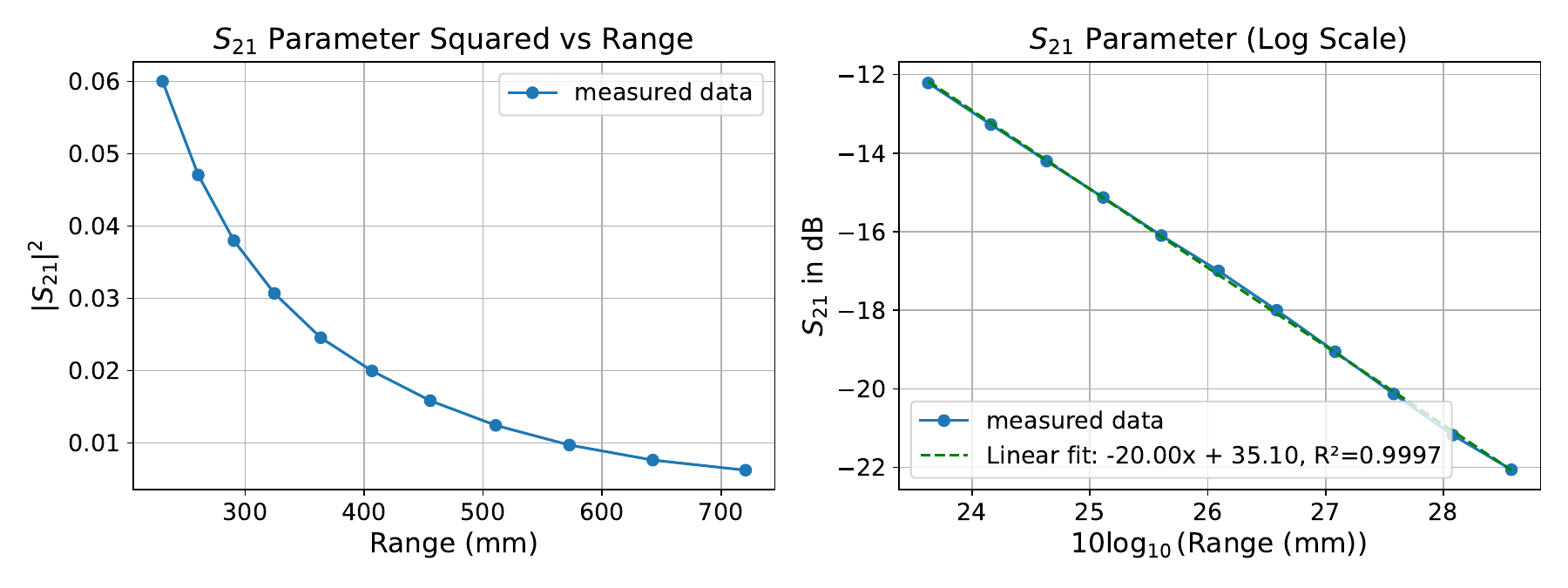}
    \caption{Antenna gain measurement experiment.}
    \label{suppfig:antenna_gain}
\end{figure}

To characterize the antenna gain, two antennas are aligned to face each other and connected to the ports of a vector network analyzer (VNA), which measures the $S_{21}$ parameter at a target frequency. We record the $S$-parameters over varying distances within the far-field region. According to the Friis transmission equation \cite{friis1946note}, at a range $r$, the relationship is given by:
\begin{equation}
    S_{21} = -20\log_{10}(r) + G_{\mathrm{TX}} + G_{\mathrm{RX}} + 20\log_{10}\left(\frac{\lambda}{4\pi}\right) \quad [\mathrm{dB}].
\end{equation}
In practice, the range $r$ is determined by measuring the physical distance between the horn apertures. We then apply a $20.6$~mm correction to account for the phase center offset relative to the horn front end. This offset value was obtained \hl{from} electromagnetic simulations \hl{using} the specific dimensions of the WR-75 waveguide. As shown in \autoref{suppfig:antenna_gain}, the measured intercept is $b = 35.10$, which yields $G_{\mathrm{TX}} + G_{\mathrm{RX}} = b - 20\log_{10}(\lambda/(4\pi)) = 28.37~\mathrm{dB}$ at 11~GHz.

\subsection{Noise Bandwidth Characterization}

\begin{figure}[!ht]
\captionsetup{font=footnotesize}
\centering
\includegraphics[width=0.5\linewidth]{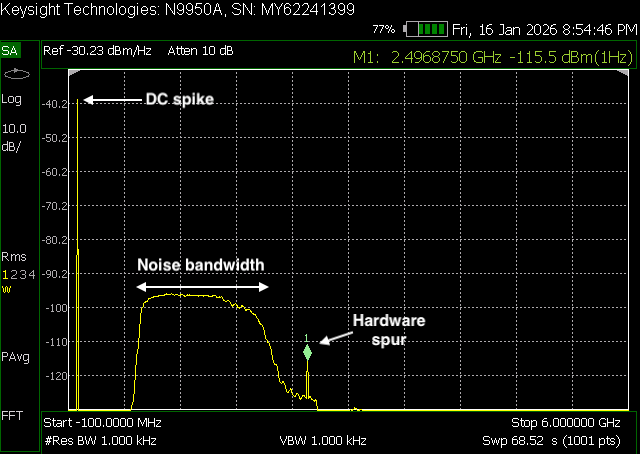}
\caption{\textbf{Power spectral density measurement of receiver front end.} The power spectral density measured at the output of the LNA with the LNB terminated is used for characterization of the system's ENBW.} 
\label{figsupp:ENBW}
\end{figure}

Two devices were used for sampling: the SDR and the ADC. The SDR's programmable bandwidth defines the system's bandwidth. For wide bandwidth experiments that use a sampling rate higher than the system's noise bandwidth, the front end's noise bandwidth must be determined empirically. The effective noise bandwidth (ENBW) of the system was characterized by measuring the system's front end. A spectrum analyzer was used to measure the power spectral density at the LNA output with the LNB input terminated. \autoref{figsupp:ENBW} shows a screenshot of the measurement. Using these points, the ENBW was estimated by integrating over the bands and using the top 20~\% of samples within the band as the reference power level. This resulted in an estimated ENBW of 982~MHz.

\section{Thermal Noise Control Experiments}
\label{suppsec:control}

\begin{figure}[!ht]
\captionsetup{font=footnotesize}
\centering
\includegraphics[width=1\linewidth]{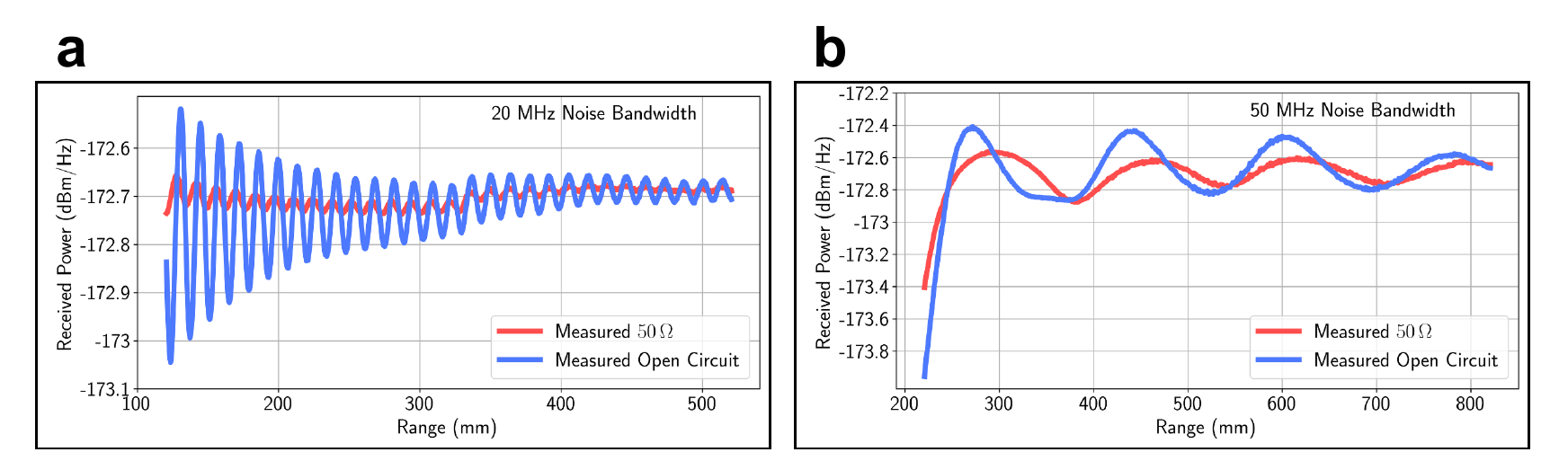}
\caption{\textbf{Thermal noise control experiments. a} Power was measured over a 50~MHz bandwidth and filtered\hl{ to use only the upper 20~MHz bandwidth at a center frequency of 11.2~GHz. The transmitter was switched between a $50~\Omega$ terminator and an open circuit. \textbf{b} A 940~MHz center-frequency measurement setup with a 50~MHz bandwidth was used to measure noise power}. The transmitter was switched between a $50~\Omega$ terminator and an open circuit.} 
\label{figsupp:control}
\end{figure}

To ensure that the measured power from our experiments was thermal noise, several control experiments were conducted. The receiver hardware was configured with a bandwidth of 50~MHz, and the transmitter was terminated in a $50~\Omega$ resistor and an open circuit. Instead of processing the samples over the full measured bandwidth, a fast Fourier transform was used to filter out the bottom 30~MHz of the measurement bandwidth. This ensured that the measured oscillations were not due to local oscillator leakage. As seen in \autoref{figsupp:control}a, the matched terminator behavior remains the same, and the power oscillations in the open-circuit case persist.

In addition, we used several hardware configurations to ensure reproducibility and avoid hardware-specific effects. In the paper, results are shown using an 11~GHz center frequency setup with two different receivers, the Ettus B210 SDR and the fast sampling ADC. We also conducted experiments using the SDR at a center frequency of 940~MHz. The $50~\Omega$ load and open circuit are again measured in \autoref{figsupp:control}b. The wavelength of the oscillations matches the operating frequency as expected. The \hl{antenna's structural mode appears stronger, as its oscillations are more evident using the $50~\Omega$ load.}

\section{Derivation of Mismatched Load LNA Noise Power}
\label{suppsec:eqplna}
The voltage signal at the LNA input at time $t$ is 
\begin{equation}
    y_{\rm LNA}(t) = x_{\rm FW}(t) + x_{\rm BK}'(t).
\end{equation}
Recall that $x_{\rm FW}(t)$ is the forward-propagating LNA noise, which is the noise added to a signal as it propagates through the LNA, referred to the LNA input, and $x_{\rm BK}'(t)$ is the backward-propagating LNA noise $x_{\rm BK}(t)$ that is backscattered from the transmitter and returns to the LNA input at time $t$,
\begin{equation}
    x_{\rm BK}'(t) = h^2\exp(\J \phi) x_{\rm BK}(t-\tau),
\end{equation}
where $h$ is the complex wireless channel gain, $\phi$ is the phase introduced by the TX load and $\tau$ is the round-trip propagation delay. The signal power contributed by the LNA is
\begin{equation}
    \begin{aligned}
        P_{\rm BK}(t)
    &= \E[|x_{\rm FW}(t)+x'_{\rm BK}(t)|^2]\\
    &= \E[|x_{\rm FW}(t)|^2]+\E[|x_{\rm BK}'(t)|^2]+2\Re\{\E[x_{\rm FW}^*(t)x_{\rm BK}'(t)]\} \\
    &= \E[|x_{\rm FW}(t)|^2]+|h|^4\E[|x_{\rm BK}(t-\tau)|^2]+2\Re\{h^2\exp(\J\phi)\E[x_{\rm FW}^*(t)x_{\rm BK}(t-\tau)]\} \\
    &= \E[|x_{\rm FW}(t)|^2]+|h|^4\E[|x_{\rm BK}(t-\tau)|^2]+2\Re\{h^2\exp(\J\phi)\}R_{X_{\rm FW}X_{\rm BK}}(\tau) \\
    \end{aligned}
\end{equation}
The cross-correlation function of two band-limited Gaussian processes $R_{X_{\rm FW}X_{\rm BK}}(\tau) = \rho \K\sqrt{T_{\rm BK}T_{\rm FW}} B \sinc(B\tau)$ where $\rho$ is the correlation coefficient of $x_{\rm BK}$ and $x_{\rm FW}$ and in this paper we assume $|\rho| = 1$ with $\angle \rho = \psi$. This means that all backward-propagating LNA noise is correlated with the forward-propagating noise. More generally, the input noise is modeled as a combination of correlated and uncorrelated components \cite{rogers2011noise}. Plugging in noise powers and wireless channel $h$ and assuming that the Gaussian process is stationary, we have
\begin{equation}
    P_{\rm LNA} = \K T_{\rm FW}B + \K T_{\rm BK}B\frac{G_{\mathrm{TX}}^2G_{\mathrm{RX}}^2\lambda^4}{(\mathrm{4\pi} r)^4} + 2\K \sqrt{T_{\rm BK}T_{\rm FW}}B\cos(\psi+\phi-\frac{4\pi r}{\lambda})\sinc(B\tau)\frac{G_{\mathrm{TX}}G_{\mathrm{RX}}\lambda^2}{(\mathrm{4\pi} r)^2}.
\end{equation}

\section{Matched Load Oscillations}
\label{suppsec:matchedloadoscillations}

\begin{figure}[!ht]
\captionsetup{font=footnotesize}
\centering
\includegraphics[width=0.6\linewidth]{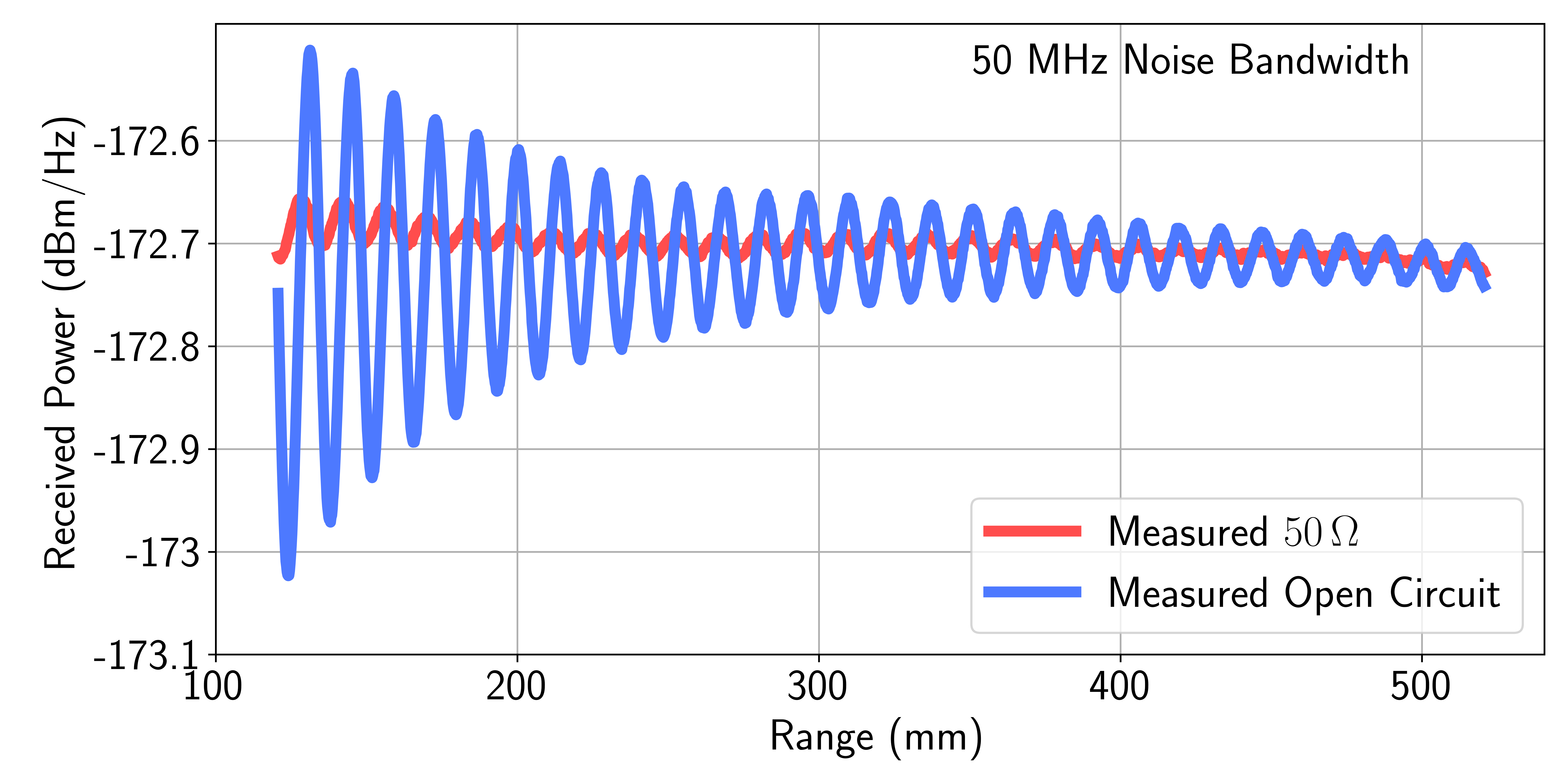}
\caption{\textbf{Measurement of matched load oscillations occurring with low noise bandwidths.} Power received when the transmitter is terminated in an open circuit and $50~\Omega$ \hl{terminator was} measured with a 50~MHz noise bandwidth.} 
\label{figsupp:50ohmOscillation}
\end{figure}

In the paper, we showed that matched loads do not exhibit power oscillations when measured with a 982~MHz noise bandwidth. However, if a low bandwidth is used, such as 50~MHz, there will be small power oscillations. This imperfection arises because the backscatter from an antenna comprises two components: antenna and structural modes. The antenna mode is due to reradiation and is controlled by the antenna's load, while the structural component is inherent to the antenna regardless of how it is matched \cite{Anderson_2005_AntennaAbsorption}. This is evident in low bandwidth measurements obtained using a 50~MHz noise bandwidth to characterize a $50~\Omega$ load and an open circuit in \autoref{figsupp:50ohmOscillation}. While the amplitude of the oscillation is much higher using the open circuit, small oscillations appear in the $50~\Omega$ measurement. This is due to the structural mode of the antenna and will depend highly on its geometry. 

\section{Channel Performance Derivation}
\label{suppsec:channel_performance}
The average power $Y$ measured at the receiver is used to detect the transmitted bit $X$. When the bandwidth $B$ is much greater than the modulation bit rate $\fb$, then in a sampling time $1/\fb$, 
$Y|X=i \sim \mathcal{N}(\mu_i, \sigma_i^2)$ where $\mu_i=P_i$ and $\sigma_i^2 = P_i^2 \fb/B$. Without loss of generality, we assume $P_1>P_0$. We assume equiprobable bits $\Pr(X=0)=\Pr(X=1)=\tfrac{1}{2}$ and that $P_1$ and $P_0$ are at least approximately equal, which holds in wireless thermal noise backscatter links. Under these conditions, the maximum likelihood decision threshold is well approximated by $P_D = \frac{\sigma_0\mu_1+\sigma_1\mu_0}{\sigma_0+\sigma_1} = \frac{2P_0P_1}{P_0+P_1}$. We define a power contrast ratio $\widetilde{\Delta P} = \frac{P_1-P_0}{P_1+P_0}$ and use it to express a Q-factor, which is a normalized measure of 
the separation between the conditional means relative to the noise standard deviations: 
\begin{align}
Q := \frac{\mu_1-\mu_0}{\sigma_1+\sigma_0}=\widetilde{\Delta P}\sqrt{\frac{B}{R_{\mathrm{b}}}},
\label{eq:Q_factor}
\end{align}
Assuming the decision threshold is set to $P_D$, the \hl{bit-error probability} is given by
\begin{equation}
\label{eq:BER}
P_{\rm b} = \tfrac{1}{2}\mathrm{erfc}\bigl(Q/\sqrt{2}\bigr)
= \tfrac{1}{2}\mathrm{erfc}\left(\pcr\sqrt{\frac{B}{2R_{\mathrm{b}}}}\right).
\end{equation}

In a wireless thermal backscatter link, $P_0$ and $P_1$ are nearly equal. Therefore, the conditional PDFs for bits 0 and 1 have nearly equal variances, and the channel is well approximated by an additive white Gaussian noise (AWGN) channel with binary input. The bit error ratio of binary-input AWGN is given by 
$P_{\rm b}=\frac{1}{2}\mathrm{erfc}(\sqrt{\mathrm{SNR}})$. Equating this to the previous bit-error rate expression Eq.~\ref{eq:BER} yields 
\begin{equation}
    {\rm SNR}=(\pcr)^2 \frac{B}{\fb}.
\end{equation}

\section{Communication System Implementation}
\label{suppsec:comm_sys_imp}
Regarding the transmitter, a Raspberry Pi Zero is utilized to control the RF switch by toggling a designated GPIO pin according to the input bit stream at the specified modulation bit rate. The input bit stream is generated from a random information bit sequence, which is subsequently processed using channel coding to enhance error resilience. On the receiver side, raw digital I/Q samples are recorded asynchronously using the USRP. The captured data is subsequently processed in MATLAB, where a signal processing pipeline performs frame synchronization and channel decoding to recover the original information bits. Finally, the decoded bit stream is compared against the transmitted sequence to determine the \hl{bit-error probability} and verify the reliability of the communication link. 

\noindent \textbf{Transmitter.} Each data frame consists of a 2047-bit preamble for synchronization,  followed by a 4096-bit payload. We employ an $m$-sequence (denoted as ${\bf p}$, with degree $11$) for the preamble due to its low implementation complexity and excellent autocorrelation properties. For the payload, we utilize Polar coding for error correction. The transmitter begins with $K$ randomly generated information bits $\mathbf{u} \in \{0,1\}^K$. For a codeword length $N = 4096$, the Polar coding algorithm identifies the frozen indices and utilizes a generator matrix $\mathbf{G} \in \{0,1\}^{N \times K}$ to produce the payload sequence $\mathbf{x} = \mathbf{G}\mathbf{u}$. Consequently, a complete data frame is represented as $\mathbf{y} = [\mathbf{p} ; \mathbf{x}]$. The code rate is defined as $R = K/N$; generally, a higher rate provides less resilience against channel errors. In our experiments, we evaluated various code rates and identified the maximum rate that permitted error-free decoding under the given channel conditions.

\noindent {\bf Receiver.} We assume the receiver is tuned to the optimal operating frequency. This is determined by a calibration phase where the transmitter continuously toggles between the two binary states while the receiver scans the available noise bandwidth. The receiver selects the center frequency that maximizes the thermal contrast between these states and maintains this frequency for subsequent data acquisition. 

The recorded complex baseband signal $x[n]$ is first converted into a power sequence $p[n] = |x[n]|^2$. This sequence is then cross-correlated with a local template of the preamble, which is upsampled to match the ratio between the receiver's sampling rate and the modulation bit rate $R_{\mathrm{b}}$. Data frame detection and synchronization are achieved by identifying the correlation peaks and their corresponding temporal indices.

For channel estimation, the receiver utilizes the known preamble sequence to estimate the average power levels corresponding to the binary 0 and 1 states. Let $\mathcal{I}_0 = \{i \mid p[i] = 0\}$ and $\mathcal{I}_1 = \{i \mid p[i] = 1\}$ denote the sets of indices for bits 0 and 1 within the preamble, respectively. The estimated power for the $j$-th bit state, $\hat{p}_j$, is calculated as:
\begin{equation}
    \hat{p}_j = \frac{1}{|\mathcal{I}_j|} \sum_{i \in \mathcal{I}_j} p[i], \quad j \in \{0, 1\}
\end{equation}
where $|\mathcal{I}_j|$ represents the cardinality of the index set. We also estimate the empirical variance of the distribution of the two states:
\begin{equation}
    \hat{\sigma}_j^2 = \frac{1}{|\mathcal{I}_j|} \sum_{i \in \mathcal{I}_j} (p[i] - \hat{p}_j)^2, \quad j \in \{0, 1\}
\end{equation}
In practice, the variances of the two noise states are approximately equal due to the similar power levels of the respective noise distributions. Consequently, we utilize an averaged empirical variance for the decoding process: $\hat{\sigma}^2 = \frac{\hat{\sigma}_0^2 + \hat{\sigma}_1^2}{2}$.

Using the data frame starting index and the estimated power levels for the binary states, the receiver performs data demodulation and decoding. For demodulation, we first average the power samples within each bit duration to obtain the vector $\mathbf{y} \in \mathbb{R}^{N}$. Assuming a Gaussian distribution for the averaged power, the log-likelihood ratio (LLR) is given by:
\begin{equation}
    \mathrm{LLR} = \ln \left( \frac{\Pr(\mathbf{y} \mid x=0)}{\Pr(\mathbf{y} \mid x=1)} \right) = \frac{1}{2\hat{\sigma}^2} \left[ (\mathbf{y} - \hat{p}_1)^2 - (\mathbf{y} - \hat{p}_0)^2 \right]
\end{equation}

Finally, the computed LLRs are fed into the Polar decoder to recover the original information bits. Our implementation utilizes CRC-aided successive-cancellation list (SCL) decoding \cite{tal2015list} with a list size of $L=1024$ to achieve near-capacity performance. For the decoding process, we leverage the open-source MATLAB implementation provided in \cite{yu2021polar}.
\putbib[refs]
\end{bibunit}

\end{document}